\documentclass[oneside,10pt,english]{swissi}
\newcommand{\papertitle}{Accountable AI Delegation in Banking}
\newcommand{\papersubtitle}{Identity, Commitment Controls, Decision Evidence and Recovery}

\newcommand{\papercitationauthors}{Kurz, W. and Alidemi, B. and Schmidt, M.}

\newcommand{\paperauthors}{%
  Walter Kurz\textsuperscript{1}%
  , Besnik Alidemi\textsuperscript{2}%
  , Marco Schmidt\textsuperscript{2}%
}

\newcommand{\paperaffiliations}{%
  \textsuperscript{1}Swissi Institute for AI,
  \href{mailto:office@swissi-ai.institute}{office@swissi-ai.institute}\\
  \textsuperscript{2}Signum Magnum College%
}

\newcommand{\paperdeclarationaffiliations}{\begin{tabular}[t]{@{}p{\linewidth}@{}}
  \textsuperscript{1}Swissi Institute for AI,
  \href{mailto:office@swissi-ai.institute}{office@swissi-ai.institute}\\
  \textsuperscript{2}Signum Magnum College%

\end{tabular}}

\newcommand{\papercorresponding}{Walter Kurz, Swissi Institute for AI, \href{mailto:office@swissi-ai.institute}{office@swissi-ai.institute}. ORCID: \href{https://orcid.org/0009-0006-8045-4775}{0009-0006-8045-4775}.}

\newcommand{\paperfunding}{This research received no funding.}
\newcommand{\paperconflictsofinterest}{The authors declare no conflicts of interest.}
\newcommand{\paperdataavailability}{Survey question wording, rating anchors and aggregate findings are provided in Supplementary Material S1. The study protocol, synthetic data, source code, expected outcomes and complete computational results accompany Supplementary Material S2. Individual survey responses are not distributed.}
\newcommand{\paperaitools}{According to guidelines. Codex (Astra 6.2) and Claude Code (Fable 5.1) were used to test the Python implementation.}
\newcommand{\paperauthorcontributions}{All authors contributed equally to this work.}

\DeclareFontFamily{T1}{swissifigure}{}
\DeclareFontShape{T1}{swissifigure}{m}{n}{<-> ecrm1000}{}
\DeclareFontShape{T1}{swissifigure}{bx}{n}{<-> ecbx1000}{}
\DeclareFontShape{T1}{swissifigure}{b}{n}{<->ssub*swissifigure/bx/n}{}
\newcommand*{\swFigureFont}{\microtypesetup{expansion=false}\fontfamily{swissifigure}\selectfont}

\DefineBibliographyStrings{english}{urlseen={Last verified}}
\AtEveryBibitem{\iffieldequalstr{journaltitle}{Swissi AI Journal}{\clearfield{doi}}{}}

\hypersetup{pdftitle={\papertitle},pdfauthor={Walter Kurz, Besnik Alidemi and Marco Schmidt},pdfsubject={Institutional delegation and synthetic control evaluation}}

\title{\papertitle \\[0.35em]
\normalsize \papersubtitle}

\author{{\fontsize{8}{10}\selectfont\paperauthors}\\[0.9em]
{\scriptsize\paperaffiliations}}
\date{\vspace{1.5em}}

\begin{document}
\maketitle
\begin{abstract}
An AI-assisted payment can pass assessment and lose release authority before commitment. We specify a banking delegation contract that checks current authority, the shared payment cap and the capacity of eligible instruments to cover losses. It books the new payment and its recovery demands and writes the outcome record in the same atomic transition. A paired example identifies the authorised mandate, applied rule and selection inputs a reviewer needs to assess mandate conformance. We show that, under fixed resource inputs, the transition preserves the payment cap and recovery feasibility when commitments and effective withdrawals are ordered against the same authoritative state. Reversing a transmitted priority changes the selected route in 4 of 32 distinct behavioural combinations, represented by 96 of 576 labelled contexts. The executor follows the transmitted rule in all of them. Reusing earlier state admits 23 releases in 123 histories blocked by the reference. Pooling recovery capacity admits 69 infeasible configurations in a grid of 576. Both current-state implementations agree with the reference in all 144 commitment histories and 624 concurrent schedules. The retention test recovers five specified outcome fields from each of 47 complete records. Fifteen survey responses informed requirements for resolving holds and handling uncertain delivery. An insurance example illustrates referral after an adverse coverage finding.

\end{abstract}
\keywords{banking; AI delegation; institutional accountability; principal-agent theory; revocation; recovery capacity; decision evidence}

\section{Introduction}

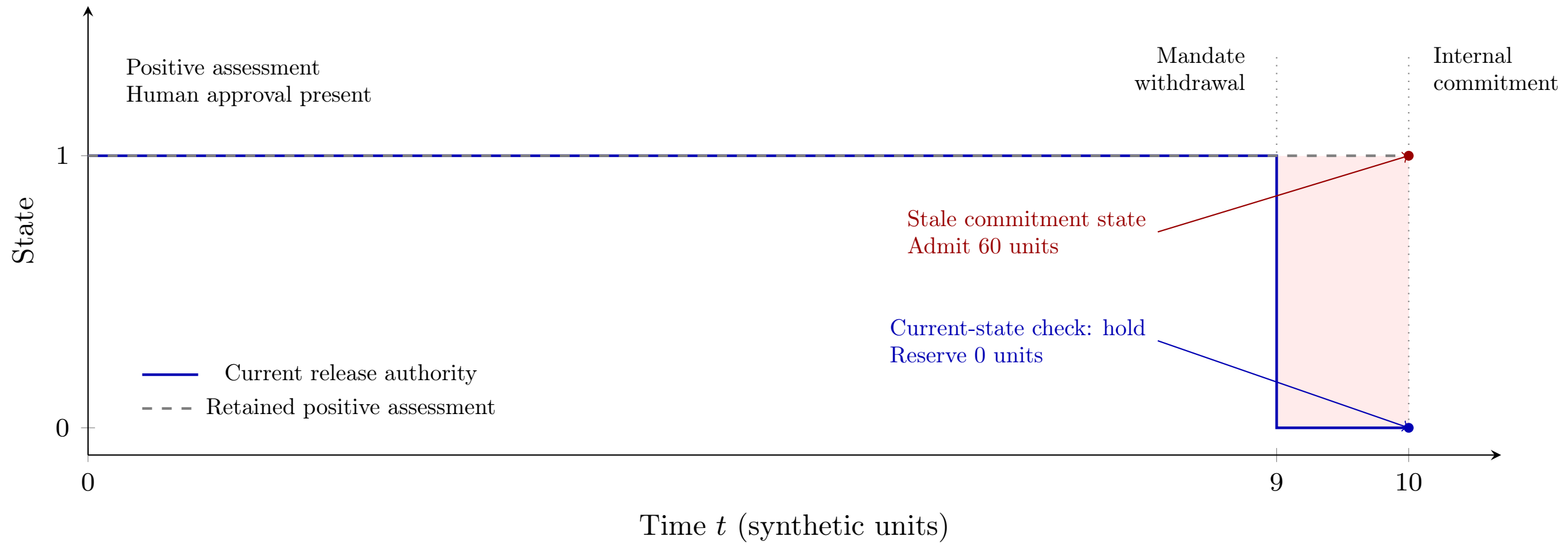
\begin{figure}[htbp]
\centering
\begin{tikzpicture}
\begin{axis}[
width=0.93\linewidth,height=6.4cm,axis lines=left,
xmin=0,xmax=10.7,ymin=-0.1,ymax=1.55,
xtick={0,9,10},ytick={0,1},
xlabel={Time $t$ (synthetic units)},ylabel={State},
tick label style={font=\footnotesize\swFigureFont},label style={font=\small\swFigureFont},
legend style={font=\scriptsize\swFigureFont,draw=none,at={(0.03,0.05)},anchor=south west},
clip=false]
\fill[red!8] (axis cs:9,0)--(axis cs:10,0)--(axis cs:10,1)--(axis cs:9,1)--cycle;
\draw[dotted,gray] (axis cs:9,0)--(axis cs:9,1.38);
\draw[dotted,gray] (axis cs:10,0)--(axis cs:10,1.38);
\addplot[thick,blue!70!black] coordinates {(0,1)(9,1)(9,0)(10,0)};
\addlegendentry{Current release authority}
\addplot[thick,gray,dashed] coordinates {(0,1)(10,1)};
\addlegendentry{Retained positive assessment}
\node[font=\scriptsize\swFigureFont,anchor=west,align=left] at (axis cs:0.2,1.27)
 {Positive assessment\\Human approval present};
\node[font=\scriptsize\swFigureFont,anchor=south east,align=right] at (axis cs:8.85,1.2)
 {Mandate\\withdrawal};
\node[font=\scriptsize\swFigureFont,anchor=south west,align=left] at (axis cs:10.1,1.2)
 {Internal\\commitment};
\node[font=\scriptsize\swFigureFont,red!60!black,align=left,anchor=east] (cachedoutcome)
 at (axis cs:8.1,0.72) {Stale commitment state\\Admit 60 units};
\draw[->,red!60!black] (cachedoutcome.east)--(axis cs:10,1);
\addplot[only marks,mark=*,mark size=1.3pt,red!60!black,forget plot]
 coordinates {(10,1)};
\node[font=\scriptsize\swFigureFont,blue!70!black,align=left,anchor=east] (liveoutcome)
 at (axis cs:8.1,0.32) {Current-state check: hold\\Reserve 0 units};
\draw[->,blue!70!black] (liveoutcome.east)--(axis cs:10,0);
\addplot[only marks,mark=*,mark size=1.3pt,blue!70!black,forget plot]
 coordinates {(10,0)};
\end{axis}
\end{tikzpicture}
\caption{Internal commitment after mandate withdrawal in the constructed payment case.}
\label{fig:case-trace}
\end{figure}

An AI-assisted payment can pass screening at assessment and lose release authority before the institution commits it. In \cref{fig:case-trace}, the earlier assessment remains correct after the mandate is withdrawn. At commitment, the current-state check holds the payment. A check that reuses the earlier state admits it. The assessment alone cannot tell a reviewer which decision the institution made or what authority it had at the time.

An executor can also follow a mistransmitted instruction exactly, and two payments that each fit the available limit at assessment can exceed it together. The institutional question is which handling rule the executor followed, whether that rule conformed to the authorised mandate, and whether the payment met the conditions for commitment. Each question requires evidence from a different point in the workflow.

The earlier wealth-management model rechecks current evidence and version-bound approval before issuing a recommendation \parencite{kurz2026nesteddelegation}. A payment admission also consumes capacity available to other requests. We extend the model with a shared payment cap and a check that eligible recovery instruments can cover outstanding and proposed losses in each declared scenario. The bank checks current authority, books the new payment and its recovery demands, and records the outcome in one internal transition. We examine this contract through a preservation argument and deterministic cases. A formative survey informed the requirements for holds and uncertain delivery.

\noindent\begin{minipage}{\linewidth}
The bank can delegate assessment and retain control over whether a proposed payment is admitted. \Cref{fig:delegation} locates the AI proposal and the employee's assigned powers in the institutional workflow. Both inform the commitment decision, which determines how the bank handles the request next.

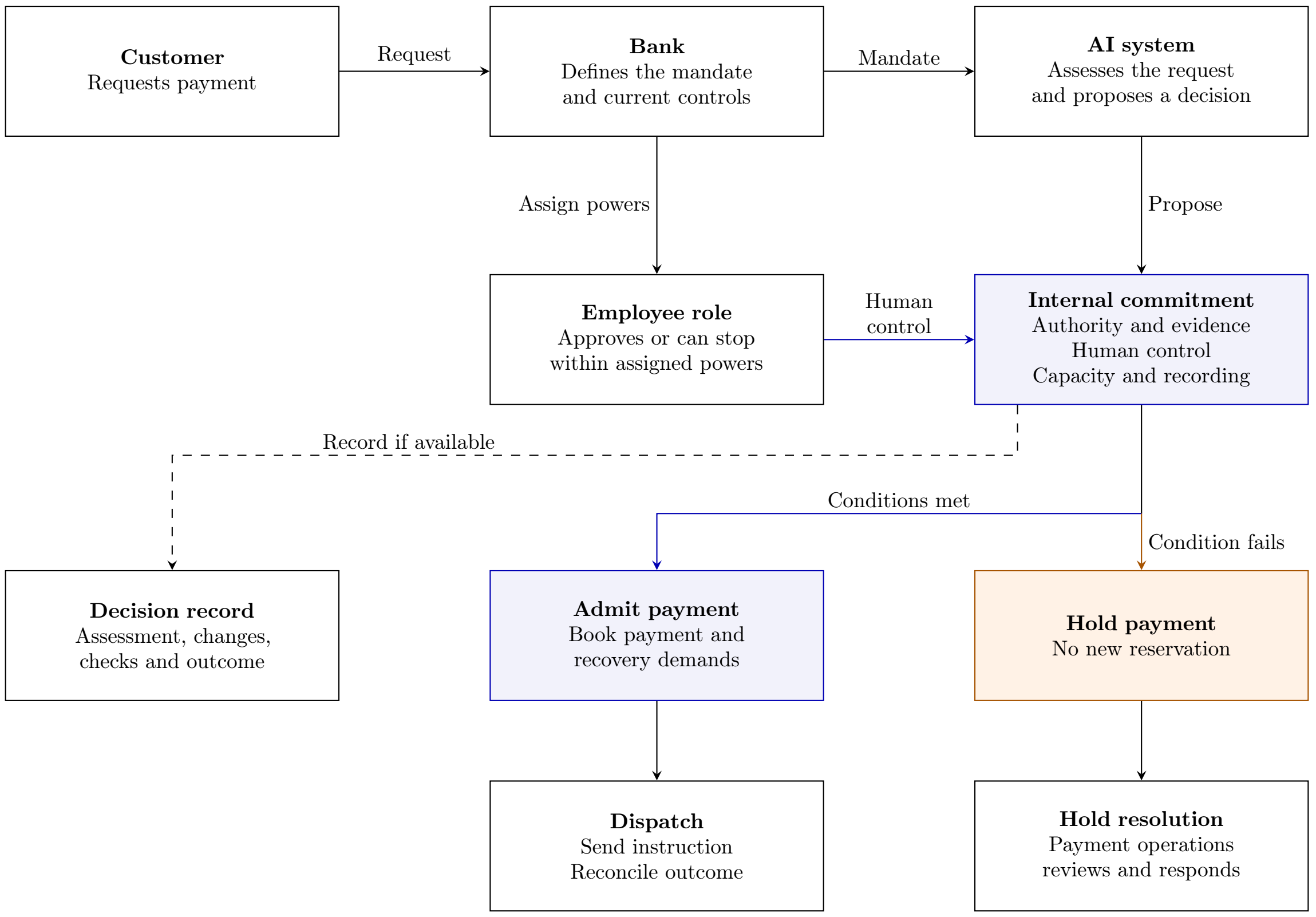
\begin{figure}[H]
\centering
\colorlet{controlBlue}{blue!70!black}
\colorlet{holdAmber}{orange!65!black}
\begin{tikzpicture}[
  >=stealth,
  box/.style={draw,thin,align=center,font=\scriptsize\swFigureFont,text width=3.5cm,
    minimum height=1.5cm,inner sep=5pt},
  flow/.style={->,thin},
  edge label/.style={font=\scriptsize\swFigureFont,align=center,inner sep=2pt}
]
  \node[box] (customer) at (0,3.1)
    {\textbf{Customer}\\Requests payment};
  \node[box] (bank) at (5.6,3.1)
    {\textbf{Bank}\\Defines the mandate\\and current controls};
  \node[box] (ai) at (11.2,3.1)
    {\textbf{AI system}\\Assesses the request\\and proposes a decision};
  \node[box] (employee) at (5.6,0)
    {\textbf{Employee role}\\Approves or can stop\\within assigned powers};
  \node[box,draw=controlBlue,fill=controlBlue!5] (gate) at (11.2,0)
    {\textbf{Internal commitment}\\Authority and evidence\\Human control\\Capacity and recording};
  \node[box] (record) at (0,-3.42)
    {\textbf{Decision record}\\Assessment, changes,\\checks and outcome};
  \node[box,draw=controlBlue,fill=controlBlue!5] (admit) at (5.6,-3.42)
    {\textbf{Admit payment}\\Book payment and\\recovery demands};
  \node[box,draw=holdAmber,fill=orange!10] (hold) at (11.2,-3.42)
    {\textbf{Hold payment}\\No new reservation};
  \node[box] (dispatch) at (5.6,-5.85)
    {\textbf{Dispatch}\\Send instruction\\Reconcile outcome};
  \node[box] (resolution) at (11.2,-5.85)
    {\textbf{Hold resolution}\\Payment operations\\reviews and responds};

  \draw[flow] (customer.east) -- node[edge label,above]{Request} (bank.west);
  \draw[flow] (bank.east) -- node[edge label,above]{Mandate} (ai.west);
  \draw[flow] (bank.south) -- node[edge label,left]{Assign powers} (employee.north);
  \draw[flow] (ai.south) -- node[edge label,right]{Propose} (gate.north);
  \draw[flow,draw=controlBlue] (employee.east) --
    node[edge label,above]{Human\\control} (gate.west);

  \draw[flow,dashed] ([xshift=5mm]gate.south west) -- ++(0,-0.58) -|
    node[edge label,pos=0.36,above]{Record if available} (record.north);
  \draw[thin] (gate.south) -- (11.2,-2.01) coordinate (outcome);
  \draw[flow,draw=controlBlue] (outcome) -|
    node[edge label,pos=0.25,above]{Conditions met} (admit.north);
  \draw[flow,draw=holdAmber] (outcome) --
    node[edge label,right]{Condition fails} (hold.north);
  \draw[flow] (admit.south) -- (dispatch.north);
  \draw[flow] (hold.south) -- (resolution.north);
\end{tikzpicture}
\caption{Proposed institutional workflow for payment delegation, commitment and subsequent handling.}
\label{fig:delegation}
\end{figure}

A held payment still requires review and a customer update. When recording is available, the outcome record gives payment operations the basis for reviewing the hold and deciding the next authorised action.
\end{minipage}
\par

\section{Related work}

Agency theory distinguishes delegated decision rights, monitoring costs and residual loss. Hierarchical models examine information transmission across successive delegations \parencite{jensen1976theory,tirole1986hierarchies}. Incomplete-contract theory addresses residual control rights when exhaustive specification is costly \parencite{grossmanhart1986}. These distinctions motivate separate responsibility for mandate specification, execution and unresolved cases. The wealth-management formulation already distinguishes errors introduced in transmission from errors in executing the transmitted instruction. It also constrains the joint objective and binds approval to the current decision version \parencite{kurz2026nesteddelegation}. The payment contract applies those controls when each admission draws on shared financial capacity.

FinaAI supplies financial policy gates and evidence requirements \parencite{kurz2025finaai}. The identity-assurance model distinguishes actor types and ties reliance to the act and time of use \parencite{kurz2026actorassurance}. Attributable ledger records preserve historical rule references \parencite{kurz2026identitystakedconsensus}.

Complete mediation requires authority checks on each access and reconsideration of cached decisions when authority changes \parencite{saltzer1975protection}. NIST separates policy decision from enforcement \parencite{nist2020zerotrust}. The Agentic Payment Protocol (AP2), version 0.2, specifies checkout and payment authority with linked dispute evidence \parencite{google2026ap2}. OAuth Token Exchange distinguishes subject and actor and leaves revocation propagation to the deployment \parencite{rfc8693}. The Continuous Access Evaluation Profile supplies revocation and claims-change signals \parencite{openid2025caep}. Financial pre-trade controls combine authorised access with aggregate exposure limits \parencite{sec2010marketaccess}, and payment authentication binds authorisation to amount and payee \parencite{ec2018dynamiclinking}.

Lan et al.\ propose binding authority to the execution context and checking a single-use nonce atomically \parencite{lan2026runtime}. Louck et al.\ propose binding spending authority to evidence of what the customer was shown \parencite{louck2026signing}. Mandato specifies signed mandates, quota checks, human escalation and chained decision records \parencite{racioppi2026mandato}. Its quota check serves a role similar to the payment cap. Our admission check also asks whether eligible instruments can cover the existing and proposed losses in each declared scenario. Liu et al.\ distinguish receipt and persistence of revocation from gateway enforcement and prevention of external effects \parencite{liu2026revocation}. We assume coherent state at internal commitment. Delays between revocation receipt and enforcement at a distributed gateway fall outside that assumption.

Linearizability gives concurrent operations an order consistent with real-time precedence \parencite{herlihywing1990}. Invariant-confluence analysis asks when independently valid transactions remain valid after their effects are merged, locating the need for coordination in the operations and constraints together \parencite{bailis2014coordination}. O'Neil's example of generalised escrow guarantees that seat reservations can be honoured while deferring the assignment of seats to passengers with no seating preference \parencite[\S~4.3]{oneil1986escrow}. At commitment, the bank checks whether eligible instruments can cover outstanding and proposed losses in every declared scenario. On admission it books those demands and can rearrange the allocations as further demands are admitted.

Reviewable automated decision-making calls for records that support review and intervention across the decision lifecycle \parencite{cobbe2021reviewable}. Here the record links the assessed disposition to the conditions and demands accepted at commitment.

\section{Institutional delegation model}

Institution $J$ delegates bounded decisions to machine $M$ and assigns approval or stopping powers to natural role $N$, following the identity-assurance taxonomy \parencite{kurz2026actorassurance}. Customer instruction, institutional control authority and runtime access have separate bases. In insurance, entitlement likewise remains distinct from delegated handling authority. \Cref{fig:delegation} connects these assignments to commitment and hold resolution. Attribution records authority and conduct; liability depends on applicable law and contractual obligations.

An act $a=(u,d,o,b,v)$ identifies request $u$, decision class $d$, object $o$, affected party $b$ and amount $v$. Mandate $m$ binds the actors, permitted acts and affected parties to a validity interval, policy $p$, configuration $\gamma$, exposure and recovery rules, and decision specification $\chi_m$. All decision-relevant references are versioned. The specification contains a finite set of handling routes $\mathcal H_m(u)$, binding conditions $\mathcal C_m$ and a reflexive, transitive preference relation $\succeq_m$. At evidence state $E_t$, selection is restricted to admissible routes:
\begin{align}
 \mathcal F_m(u,E_t)&=\{\eta\in\mathcal H_m(u):
 C(\eta;u,E_t)=1\ \forall C\in\mathcal C_m\},\nonumber\\
 \eta^*&\in\operatorname{Max}_{\succeq_m}\mathcal F_m(u,E_t).
 \label{eq:selection}
\end{align}
The mandate resolves tied or incomparable maximal routes and assigns an exception process for an empty feasible set. Routes obtain evidence, clarification or review, or proceed with a supported disposition. Binding duties constrain selection before customer consequences and institutional demands rank the remaining alternatives \parencite{kurz2026nesteddelegation}.

Let $\chi^0$ denote the authorised specification and $\chi^T=T_J(\chi^0)$ its institutional transmission. For a fixed request, evidence state and tie rule, execution fidelity compares $\operatorname{Execute}(\chi^T,u,E_t)$ with $\operatorname{Select}(\chi^T,u,E_t)$. Mandate conformance compares the executed route with $\operatorname{Select}(\chi^0,u,E_t)$ and checks the original binding conditions. Both comparisons take the institution's authorised specification as given.

Consider a 60-unit payment with an unresolved premise. A fast check takes two delay units and four effort units. Clarification takes six delay units and one effort unit. With both routes available and a deadline of eight, a customer-first mandate selects the fast check and an institution-first mandate selects clarification. Consider two histories in which the executor receives an institution-first rule and selects clarification. In the first, the institution authorised that priority before assessment. In the second, transmission changed an authorised customer-first priority. In both, clarification supplies positive evidence at time~6, approval is renewed at time~7 and the payment is committed at time~10 under current authority and with sufficient recovery capacity. A transaction record containing the amount, commitment time and disposition is identical in the two histories. Given the request, selection inputs and applied rule, a reviewer can establish fidelity in both. To decide mandate conformance, the reviewer also needs the authorised priority in force at assessment and evidence of any approved change.

Each consequential act must also satisfy current commitment conditions. Identity condition $I$ requires the act-specific confirmations for the institution, appointed person and attached machine. Scope $S_m$ checks delegated powers, affected parties, validity and limits; $P$ checks the disposition against policy; $Z$ checks required premise support and the treatment of gaps. Unresolved material premises require clarification, hold or referral as specified. Evidence retains its source and epistemic status across handoffs. Configuration condition $C_m$ checks the mandate-bound model, tools, prompts and governing rules; $R_m$ denotes effective revocation. Required screening, its validity and the changes demanding renewed checks are service-specific.

Human control $H$ applies the mandate's assigned mode $\omega_m(a)$. Autonomous permission is $U_m$; approval token $z$ must be current and bound to the act, mandate, policy, configuration and decision state $s_t$. For monitoring, $V_N$ requires current stopping power, available staff and sufficient information. With event time $t_e$, commitment time $T$, and total detection, notification, assessment and stopping latency $\tau=\tau_{\mathrm d}+\tau_{\mathrm n}+\tau_{\mathrm a}+\tau_{\mathrm s}$,
\begin{equation}
 H(a,t)=
 \begin{cases}
 U_m(a,t), & \omega_m(a)=\mathrm{automatic},\\
 \operatorname{valid}(z,t)\wedge\operatorname{bind}(z,a,m,p,\gamma,s_t),
 & \omega_m(a)=\mathrm{approval},\\
 V_N(a,t)\wedge[\tau\le T-t_e], & \omega_m(a)=\mathrm{monitoring}.
 \end{cases}
 \label{eq:human}
\end{equation}
At equality, a completed stop must precede commitment in the event order; timing uncertainty requires a margin. Material evidence or authority changes require renewed approval. Shared consumption is rechecked separately from the approval-bound decision version. Customer confirmation supplies transaction-intent evidence where required.

Recovery condition $K_m$ requires an eligible allocation for existing and proposed exposures in every declared scenario $w$. Let $L_i(w,t)$ be exposure $i$'s loss, $C_j(w,t)$ the capacity of instrument $j$, and $q_{ij}(w,t)$ its coverage and reachability indicator. For each $w$, admission requires
\begin{align}
 \exists x_{ij}(w,t)\ge0:\quad
 &\sum_jx_{ij}(w,t)\ge L_i(w,t)\quad\forall i,\nonumber\\
 &\sum_ix_{ij}(w,t)\le C_j(w,t)\quad\forall j,\nonumber\\
 &x_{ij}(w,t)=0\quad\text{if }q_{ij}(w,t)=0.
 \label{eq:recovery}
\end{align}
Capacity is net of obligations outside the included exposure set. Included obligations enter once through their loss demands. Each scenario may have a different allocation. Losses that can arise together must appear together in a scenario, and its allocation must be feasible within the instruments' access and payout windows. The Basel Committee's guidance on bank liquidity risk calls for testing institution-specific and market-wide stresses separately and in combination \parencite[Principle~10]{bcbs2008liquidity}. The recovery specification supplies the access and payout windows, beneficiaries, exclusions and priority. The test takes the specified losses and eligibility as inputs.

Scenario-specific allocations assume the institution can distinguish the scenarios before the relevant allocations become binding. Suppose one instrument can cover either of two beneficiaries but must name one before it is known which will incur a loss. Each scenario can be feasible on its own while neither beneficiary choice covers both scenarios. For instruments with such deadlines, the admission test must also require that allocations already due at a deadline agree across scenarios the institution cannot yet distinguish.

With recording availability $W$, the gate combines these conditions:
\begin{equation}
 \begin{aligned}
 A_m(a,t)={}&I(a,t)\wedge S_m(a,t)\wedge P(a,t)\wedge Z(a,t)\\
 &\wedge C_m(a,t)\wedge H(a,t)\wedge\neg R_m(t)\wedge K_m(a,t)\wedge W(a,t).
 \end{aligned}
 \label{eq:gate}
\end{equation}
A proposed release requires positive support at internal commitment. A documented gap can justify an authorised hold. Refusal and referral have their own scope and evidence requirements.

Commitment evaluates the gate against coherent authoritative state $s$, including its clock, and books demands in the same transition as the outcome record. Let $B$ be the payment cap, $c$ prior consumption, $\mathcal L_w(s)$ the admitted loss map, $e_a$ the outcome record and $\mathcal E(s')$ the retained records in the resulting state. A successful commitment satisfies
\begin{align}
 c'&=c+v\le B,\nonumber\\
 \mathcal L_w(s')&=\mathcal L_w(s)\cup\{a:L_a(w,s)\}\quad\forall w,\nonumber\\
 \operatorname{commit}(a,s)&\Rightarrow[A_m(a,s)=1\wedge e_a\in\mathcal E(s')].
 \label{eq:commit}
\end{align}
Every new exposure has a unique identifier. Recovery admission checks existing plus proposed losses; booking retains those demands and permits eligible allocation to be rearranged. Payment consumption, admitted losses and the outcome record share one atomic transition. Commitment and effective invalidation operations are assumed linearizable in a common order that respects real-time precedence of completed operations \parencite{herlihywing1990}.

\noindent\begin{minipage}{\linewidth}
A favourable assessment supports a release proposal. Before commitment, authority may be withdrawn, evidence may expire or another payment may consume shared capacity. \Cref{fig:assessment-commitment} shows how the gate in \eqref{eq:gate} determines whether the proposal can be admitted under the conditions at commitment.

\begin{figure}[H]
\centering
\begin{tikzpicture}[
  >=stealth,
  box/.style={draw,thin,align=center,font=\scriptsize\swFigureFont,text width=3.8cm,
    minimum height=1.45cm,inner sep=5pt},
  connector/.style={thin},
  flow/.style={connector,->},
  edge label/.style={font=\scriptsize\swFigureFont,align=center,inner sep=1.5pt}
]
  \node[box] (assessment) at (0,0)
    {\textbf{Assessment}\\Apply the mandate\\and assess evidence};
  \node[box] (commitment) at (5.5,0)
    {\textbf{Internal commitment}\\Recheck conditions\\for the proposed release};
  \node[box] (conditions) at (5.5,2.5)
    {\textbf{Current conditions}\\Authority and evidence\\Human control\\Capacity and recording};
  \node[box] (admit) at (11,1)
    {\textbf{Admit payment}\\Reserve capacity\\and record the outcome};
  \node[box] (hold) at (11,-1)
    {\textbf{Hold payment}\\No new reservation\\Record reason if possible};

  \draw[flow] (assessment.east) --
    node[edge label,above]{Proposed\\release} (commitment.west);
  \draw[flow,blue!70!black] (conditions.south) -- (commitment.north);
  \draw[connector] (commitment.east) -- ++(0.65,0) coordinate (branch);
  \draw[flow] (branch) |- node[edge label,near end,above]{Pass} (admit.west);
  \draw[flow] (branch) |- node[edge label,near end,above]{Fail} (hold.west);
\end{tikzpicture}
\caption{Internal commitment of a proposed payment release.}
\label{fig:assessment-commitment}
\end{figure}
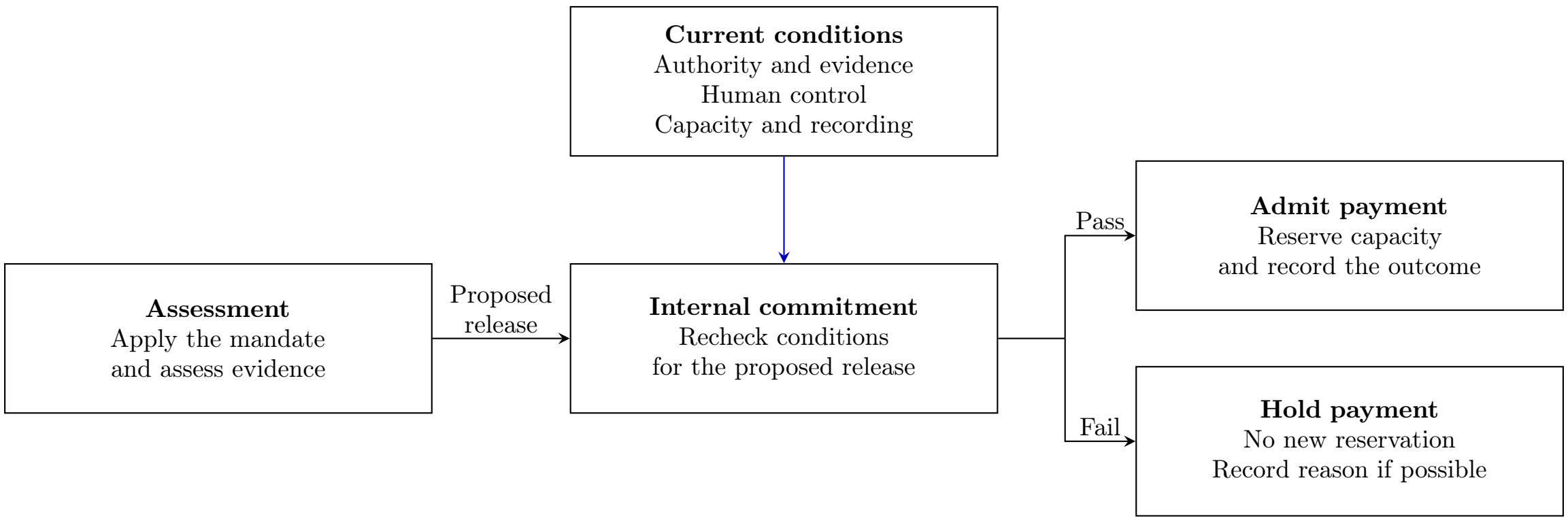

A failed check leaves the proposed payment on hold without a new reservation. When recording is available, the outcome record retains a reason and evidence reference for each check. On admission, checking capacity and booking demands in the same transition prevents a competing payment from consuming that capacity between those operations.
\end{minipage}
\par

\begin{proposition}
Assume that every act under the contract passes through the internal commitment gate. Commitment and effective invalidation are linearizable against the same authoritative state in an order that respects real-time precedence. Gate evaluation, demand booking and record creation are atomic within each commitment. Every successful commitment satisfying \eqref{eq:gate} and \eqref{eq:commit} has current authority and an associated outcome record. An effective withdrawal ordered before commitment prevents admission under that authority. Initially valid payment bounds remain satisfied under fixed caps and non-negative amounts. Recovery feasibility is preserved across admissions from a feasible initial state under fixed losses, capacities and eligibility, persistent admitted demands and permitted allocation rearrangement.
\end{proposition}
\begin{proof}
Order commitments and effective invalidations by their linearization points. Each gate sees prior commitments and invalidations, including any effective withdrawal. The payment update preserves the cap by induction. Recovery admission supplies an allocation for existing and proposed losses in each scenario. Booking retains that demand set, so the same allocations witness feasibility in the resulting state under the fixed assumptions. The atomic update writes an outcome record for each successful booking.
\end{proof}

The contract calls for an outcome record linking the request and actors to the authorised and applied specifications, the policy and configuration versions, and the assessed disposition. The record identifies the alternatives and evidence available at selection and gives the reason for the selected route. At commitment it retains the gate reasons, the order of relevant changes, and the resources reserved alongside the release or hold decision. Where the effective time of an event differs from its record time, both are needed to reconstruct the state. Evidence references must resolve to retrievable content. Signature material and custody and access rules govern how that evidence can be checked and used.

Internal commitment admits the payment and books its demands. Dispatch is the subsequent attempt to deliver the instruction. The institution must require durable pre-dispatch evidence across queued, automated and manual paths and link dispatch, acknowledgement and reconciliation to the same request. A late acknowledgement may refer to a valid commitment made before withdrawal. Diagnosing a release after withdrawal therefore requires the order of withdrawal and commitment as well as the dispatch and effect evidence. The service policy must state whether pending dispatch requires a fresh authority check. Stable request identities and bound idempotency keys must support single consumption. If delivery is uncertain, the institution retains the exposure until reconciliation determines whether redispatch or release of the reservation is permitted. Capacity or eligibility changes require reassessment of existing obligations and of further admissions.

A hold assigns a responsible role and delegated replacement, a permitted next action, review and customer-update deadlines, and cutoff-aware escalation. Resolution may obtain evidence, renew approval, refuse the instruction or propose release through the same gate. The mandate specifies any independent review required for sensitive cases. A timeout escalates the case without granting release authority or erasing an outstanding exposure. These operational requirements extend the evaluated transition to unresolved service states.

\section{Methods}

We evaluated the specification through deterministic finite designs derived from a practitioner-informed payment workflow. The rules, routes, loss scenarios and timing are constructed inputs. At assessment, verified favourable evidence supports an approval disposition and verified adverse evidence requires refusal. Missing, disputed or assumed support requires resolution. In the claims illustration, an adverse coverage finding goes to referral. Acting on or communicating any disposition requires the corresponding act-specific authority.

The baseline uses a 60-unit payment, prior consumption of 20 and cap of 100. An autonomous threshold of 50 assigns this payment to approval mode. Assessment occurs at time zero and commitment at time 10; same-time invalidations precede commitment. Synthetic units and clocks isolate ordering and arithmetic boundaries. \Cref{tab:requirements} identifies what the experiments test and what remains an implementation or domain assumption.

\begin{table}[htbp]
\centering\small
\caption{Evaluation coverage and assumptions.}
\label{tab:requirements}
\begin{tabularx}{\textwidth}{@{}L{3.2cm}>{\raggedright\arraybackslash}X>{\raggedright\arraybackslash}X@{}}
\toprule
Requirement & Evaluated mechanism & Outside the evaluation \\
\midrule
Decision specification & Ordinal selection; transmission changes & Mandate adequacy; calibrated preferences \\
Current authority and support & Ordered invalidation; evidence completion and expiry & Identity assurance; external source validity \\
Human control & Version-bound approval; declared monitoring budgets & Token authenticity; staff performance \\
Shared financial capacity & Payment booking; eligible loss allocation & Loss calibration; legal recovery effectiveness \\
Outcome evidence & Five-target record reconstruction & Durable capture; custody; reviewer interpretation \\
Operational resolution & Service requirements only & Staff workflow; distributed enforcement; external reconciliation \\
\bottomrule
\end{tabularx}
\end{table}

Four handling routes have delay/effort pairs $(0,0)$ for direct handling, $(2,4)$ for fast checking, $(6,1)$ for clarification followed by an institutional check, and $(4,3)$ for specialist review. Admissibility requires availability and compliance with the delay bound. Direct handling requires verified support; evidence-seeking routes apply to unresolved premises. An established finding receives direct handling or an exception if that route is unavailable. Customer-first selection lexicographically minimises delay, effort and a fixed tie rank; institution-first selection reverses the first two priorities. These are stipulated evaluations of service consequences.

The selection design crosses two workflow labels, six premise states, two priorities, four availability patterns, delay limits of three and eight, and three future evidence answers: 576 labelled contexts. Availability permits all routes, excludes fast checking, permits only specialist review, or permits none. Verified positive and negative premises form one applicability class; four unresolved states form another. The design contains 32 distinct behavioural combinations; workflow labels and future answers repeat them to test that selection ignores unavailable future information.

Four transmission conditions preserve the original mandate, reverse its priority, omit the delay bound, or omit direct handling's support requirement. A predicate-based selector, an equivalent procedural selector and an unconstrained ablation yield 6,912 selection rows. The reference separately evaluates the original and transmitted specifications. The unconstrained executor ranks routes without filtering for delay or support. Route departures and constraint breaches are scored against both specifications.

Eight workflow contexts crossed with 18 perturbations give 144 commitment histories per executor. Contexts cover supported payments and claims, unresolved premises with positive, adverse or assumed answers, and established adverse findings. Perturbations change authority, supervisory power, configuration, support validity, recording, payment capacity or recovery eligibility; others vary approval, monitoring and route completion. Withdrawals at times nine, 10 and 11 test the commitment boundary.

\noindent\begin{minipage}{\linewidth}
The Python evaluation holds the case inputs fixed when comparing execution methods. \Cref{fig:python-evaluation} traces the comparison with the separately implemented reference and shows how the literal cases supply records for the retention test.

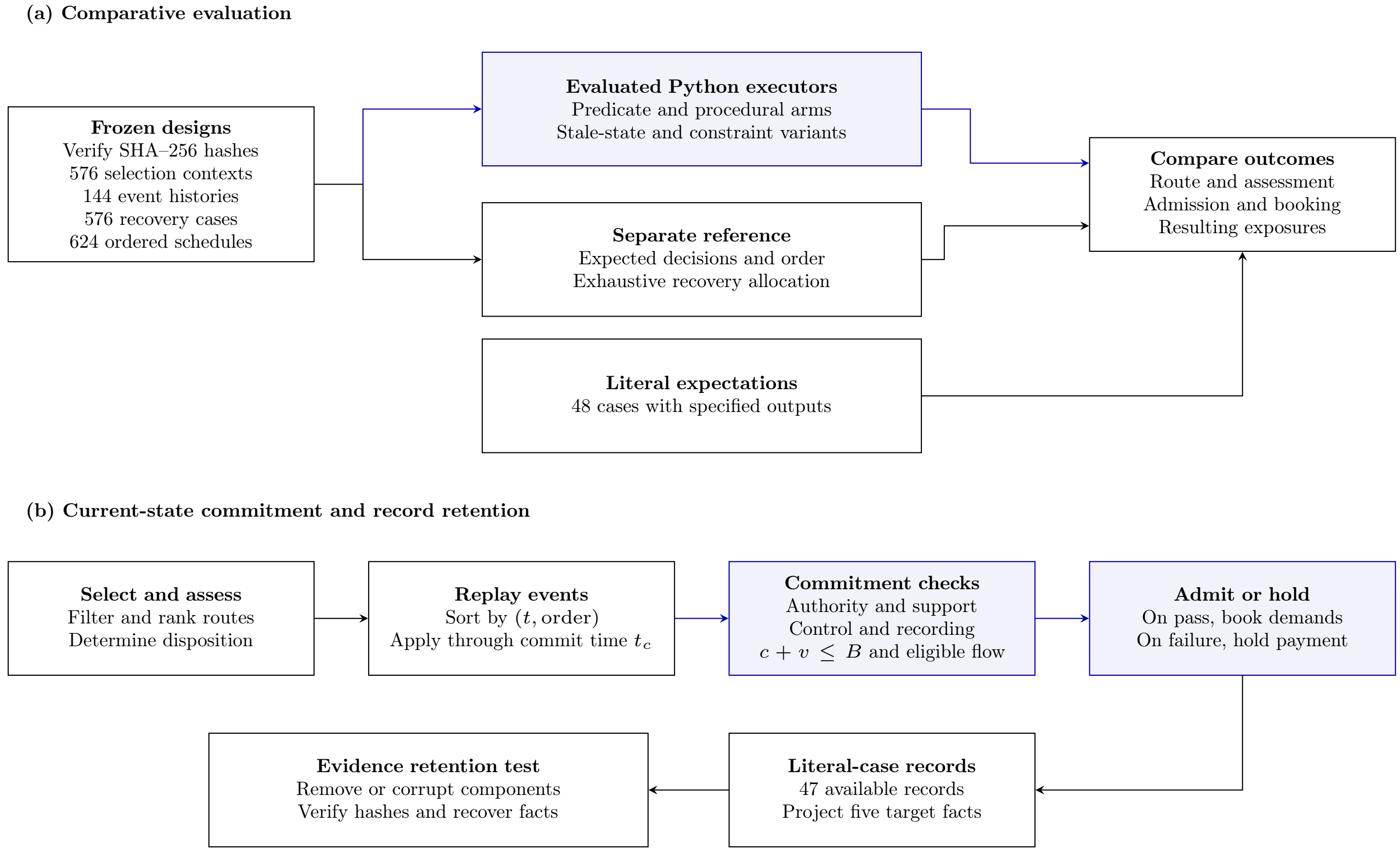
\begin{figure}[H]
\centering
\colorlet{executionBlue}{blue!70!black}
\begin{tikzpicture}[
  x=0.95cm,
  >=stealth,
  box/.style={draw,thin,align=center,font=\scriptsize\swFigureFont,text width=3.55cm,
    minimum height=1.45cm,inner sep=5pt},
  wide box/.style={box,text width=5.25cm},
  execution/.style={draw=executionBlue,fill=executionBlue!5},
  flow/.style={->,thin},
  edge label/.style={font=\scriptsize\swFigureFont,align=center,inner sep=2pt}
]
  \node[anchor=west,font=\scriptsize\swFigureFont\bfseries] at (-1.95,7.35)
    {(a) Comparative evaluation};
  \node[box] (inputs) at (0,5.18)
    {\textbf{Frozen designs}\\Verify SHA--256 hashes\\
     576 selection contexts\\144 event histories\\
     576 recovery cases\\624 ordered schedules};
  \node[wide box,execution] (executors) at (7.26,6.14)
    {\textbf{Evaluated Python executors}\\Predicate and procedural arms\\
     Stale-state and constraint variants};
  \node[wide box] (reference) at (7.26,4.22)
    {\textbf{Separate reference}\\Expected decisions and order\\
     Exhaustive recovery allocation};
  \node[wide box] (literal) at (7.26,2.48)
    {\textbf{Literal expectations}\\48 cases with specified outputs};
  \node[box] (compare) at (14.52,5.05)
    {\textbf{Compare outcomes}\\Route and assessment\\
     Admission and booking\\Resulting exposures};

  \draw[thin] (inputs.east) -- ++(0.65,0) coordinate (cases);
  \draw[flow,draw=executionBlue] (cases) |- (executors.west);
  \draw[flow] (cases) |- (reference.west);
  \draw[flow,draw=executionBlue] (executors.east) -- ++(0.65,0) |-
    ([yshift=4mm]compare.west);
  \draw[flow] (reference.east) -- ++(0.30,0) |-
    ([yshift=-4mm]compare.west);
  \draw[flow] (literal.east) -| (compare.south);

  \node[anchor=west,font=\scriptsize\swFigureFont\bfseries] at (-1.95,1.0)
    {(b) Current-state commitment and record retention};
  \node[box] (selection) at (0,-0.36)
    {\textbf{Select and assess}\\Filter and rank routes\\Determine disposition};
  \node[box] (replay) at (4.84,-0.36)
    {\textbf{Replay events}\\Sort by $(t,\mathrm{order})$\\
     Apply through commit time $t_c$};
  \node[box,execution] (checks) at (9.68,-0.36)
    {\textbf{Commitment checks}\\Authority and support\\
     Control and recording\\$c+v\le B$ and eligible flow};
  \node[box,execution] (outcome) at (14.52,-0.36)
    {\textbf{Admit or hold}\\On pass, book demands\\
     On failure, hold payment};
  \node[box] (record) at (9.68,-2.55)
    {\textbf{Literal-case records}\\47 available records\\
     Project five target facts};
  \node[wide box] (retention) at (3.59,-2.55)
    {\textbf{Evidence retention test}\\Remove or corrupt components\\
     Verify hashes and recover facts};

  \draw[flow] (selection.east) -- (replay.west);
  \draw[flow,draw=executionBlue] (replay.east) -- (checks.west);
  \draw[flow,draw=executionBlue] (checks.east) -- (outcome.west);
  \draw[flow] (outcome.south) |- (record.east);
  \draw[flow] (record.west) -- (retention.east);
\end{tikzpicture}
\caption{Python evaluation of the commitment rules. Panel (a) shows the comparative harness. Panel (b) shows current-state execution for a proposed release and reconstruction from literal-case records.}
\label{fig:python-evaluation}
\end{figure}

Scoring the stages separately can identify a route error even when final admission agrees with the reference. The retention test checks the evidence available for each fact. A missing component or a failed integrity check makes the facts supported by that component unavailable.
\end{minipage}
\par

Evidence acquisition reveals the authored answer at completion and advances the decision version before explicit same-time updates. Assessment remains held until completion. Approval tokens bind request, mandate, policy, amount, configuration and decision version, with an expiry; earlier approval becomes stale after a material change. Both live executors recheck current conditions. The stale-state ablation uses the snapshot at route completion with the planned commitment clock; unconstrained selection retains live commitment checks. Monitoring budgets of three, four and five units are compared with a four-unit intervention window.

Recovery uses normal/stress losses of $(2,4)$ for existing exposure and $(4,6)$ for the proposed exposure. General capacity six covers either; restricted capacity four covers only existing loss. The engine links each instrument to the exposures it can cover and uses breadth-first augmenting paths \parencite[\S~1.2]{edmondskarp1972} to test whether the flow covers every loss. The reference independently enumerates allocations. A 576-configuration grid varies existing and proposed losses over $\{0,2,4\}$, both capacities over $\{0,2,4,6\}$ and four restricted-instrument eligibility sets. The pooled ablation compares total loss with total capacity while ignoring eligibility.

Concurrent admission uses two 40-unit proposals, zero initial payment consumption, payment caps of 70 or 100, and general recovery capacities of eight or 12 with restricted capacity four. Losses follow the baseline construction. Six causal orders without an event and 30 with each of five events, across four capacity settings, give 624 schedules. Each check precedes its commitment; all operations share time 10 with explicit order. Events withdraw authority, expire support, disable recording, reduce the payment cap to 30 or reduce general recovery capacity to zero. The reference recomputes admission and retains earlier demands. Resource caching refreshes other gates; whole-gate caching retains the complete earlier result. Concurrency uses automatic mode; approval chronology is tested in single-request histories.

Forty-eight literal cases prescribe selected route, assessment, release, final routing label and payment reservation. Both live executors and the reference are compared with these expectations.
We used Codex (Astra 6.2) and Claude Code (Fable 5.1) to test whether the Python implementation reproduced the specified outcomes for route selection, assessment, release decisions, final routing and payment reservations. Forty-seven cases produce a record; unavailable recording produces hold without a record. Retention tests project full-method records into selection, assessment and commitment components contributing one, one and three targets. Conditions remove each component, corrupt commitment bytes under their original hash, or remove all content. Known absence is a valid answer; unavailable evidence has a separate marker.

Unauthorised release is counted among reference-blocked histories and erroneous blocking among reference-permitted histories. Selection, assessment, final routing and recovery booking receive separate error counts. Concurrency scores differences in accepted-proposal sequences separately from terminal payment excess and recovery infeasibility. A later capacity reduction can produce a shortfall after valid admission. Retention scores correct targets, complete five-target records, recognised unavailable targets and false gaps.

Supplementary Material~S2 provides the computational materials and reproduction procedure. Checksums verify that the protocol, code, test cases and inputs match the evaluated version before execution. Reported aggregates are independently recalculated from the complete computational results. The reference imports no evaluated decision code. Both live executors share disclosed state, ranking, allocation and record routines. The survey-informed operational refinements leave the evaluated implementation unchanged.

An online survey about a held payment yielded 15 responses. In the scenario, the bank withdraws the agent's release permission after detecting unusual activity. Payment operations reviews the resulting hold and decides the next authorised step. We specified the descriptive analysis after collection. We summarised each of the eight anchored 0--100 items separately and counted requests for fuller explanations. We used the comments to develop the requirements for hold resolution and delivery. Supplementary Material~S1 gives the available question wording, rating anchors, distributions and changes made in response to the feedback.

\section{Results}

The predicate-based and procedural selectors match the original and transmitted route in all 576 faithful-transmission contexts. Priority reversal changes 96 choices while preserving execution fidelity. Omitting the delay or direct-support requirement produces 120 and 192 original-constraint breaches, respectively (\cref{tab:selection-results}). Under faithful transmission, unconstrained selection chooses an inadmissible route in 240 contexts.

\begin{table}[htbp]
\centering\small
\caption{Transmission outcomes, 576 contexts per condition and selector.}
\label{tab:selection-results}
\begin{tabularx}{\textwidth}{@{}L{4.2cm}>{\raggedleft\arraybackslash}X>{\raggedleft\arraybackslash}X>{\raggedleft\arraybackslash}X@{}}
\toprule
Transmitted specification & Original route departures & Transmitted route departures & Original constraint breaches \\
\midrule
Faithful transmission & 0 & 0 & 0 \\
Priority reversed & 96 & 0 & 0 \\
Delay constraint omitted & 120 & 0 & 120 \\
Direct support omitted & 192 & 0 & 192 \\
\bottomrule
\end{tabularx}

\end{table}

The reference permits release in 21 commitment histories and blocks it in 123. Both executors that recheck current state match all reference outcomes, recovery bookings and exposure maps. Each produces 136 records, with recording unavailable in eight histories. Stale commitment state produces 23 unauthorised releases and one erroneous block. With unconstrained selection, assessment errors and three erroneous blocks occur even though the current-state check prevents unauthorised release (\cref{tab:results}). All 48 literal cases match the two executors and the reference, including cases of stale and renewed approval after evidence changes.

\begin{table}[htbp]
\centering\small
\caption{Commitment outcomes across 144 histories per executor.}
\label{tab:results}
\begin{tabularx}{\textwidth}{@{}L{4.0cm}>{\raggedleft\arraybackslash}X>{\raggedleft\arraybackslash}X>{\raggedleft\arraybackslash}X>{\raggedleft\arraybackslash}X>{\raggedleft\arraybackslash}X@{}}
\toprule
Executor & Selection errors & Assessment errors & Unauthorised releases & Erroneous blocks & Final-route errors \\
\midrule
Full controls & 0 & 0 & 0 & 0 & 0 \\
Equivalent procedural controls & 0 & 0 & 0 & 0 & 0 \\
Stale commitment state & 0 & 0 & 23 & 1 & 24 \\
Unconstrained selection & 51 & 34 & 0 & 3 & 34 \\
\bottomrule
\end{tabularx}

\end{table}

Exact eligible flow agrees with exhaustive allocation in all 576 recovery configurations. Pooling capacity admits 69 infeasible configurations and produces no false rejection.

Both implementations with current-state admission match the reference sequence in all 624 schedules. Resource caching produces 188 mismatches and whole-gate caching produces 362. Each differs from the reference in 12 of the 24 event-free schedules. Capacity reductions leave terminal payment excess in 56 schedules and recovery infeasibility in 56 even with current-state admission. \Cref{tab:concurrency-results} separates admission errors from terminal shortfalls.

\begin{table}[htbp]
\centering\small
\caption{Admission and terminal state across 624 schedules per executor.}
\label{tab:concurrency-results}
\begin{tabularx}{\textwidth}{@{}L{4.2cm}>{\raggedleft\arraybackslash}X>{\raggedleft\arraybackslash}X>{\raggedleft\arraybackslash}X@{}}
\toprule
Executor & Admission mismatches & Terminal payment excess & Terminal recovery infeasibility \\
\midrule
Full controls & 0 & 56 & 56 \\
Equivalent procedural controls & 0 & 56 & 56 \\
Cached resource checks & 188 & 152 & 152 \\
Cached whole gate & 362 & 200 & 200 \\
\bottomrule
\end{tabularx}

\end{table}

Complete retention recovers all 235 targets from 47 records. Every removal or corruption condition leaves zero complete five-target records. Expected unavailable targets are all recognised, with zero false gaps (\cref{tab:retention-results}).

\begin{table}[htbp]
\centering\small
\caption{Five-target reconstruction across 47 generated records.}
\label{tab:retention-results}
\begin{tabularx}{\textwidth}{@{}L{4.2cm}>{\raggedleft\arraybackslash}X>{\raggedleft\arraybackslash}X>{\raggedleft\arraybackslash}X@{}}
\toprule
Retention condition & Correct targets / 235 & Complete records / 47 & Recognised unavailable targets \\
\midrule
Complete retention & 235 & 47 & 0 \\
Selection absent & 188 & 0 & 47 \\
Assessment absent & 188 & 0 & 47 \\
Commitment absent & 94 & 0 & 141 \\
Commitment corrupted & 94 & 0 & 141 \\
All contents absent & 0 & 0 & 235 \\
\bottomrule
\end{tabularx}

\end{table}

The online survey returned median usefulness and withdrawal-control suitability ratings of 90/100, expected staff ease of 72 and required implementation change of 65. All 15 responses requested fuller explanation of shared resources, loss coverage and customer communication or deadlines; 12 requested more detail on recording failures. These findings informed the hold-resolution and reservation-lifecycle requirements. Nine responses omitted current-practice reporting. Full distributions and the disposition of comments appear in Supplementary Material~S1.

\section{Discussion and limitations}

The priority-reversal cases show why a correct commitment decision is insufficient evidence of mandate conformance. A reviewer can reproduce the selected route using the applied rule even if the authorised rule would have selected another. In the paired payment example, the transaction fields alone leave mandate conformance indeterminate. The decisive evidence is the authorised priority in force at assessment, linked to the applied rule and the inputs used for selection. The same evidence suffices if a procedural implementation retains it. The example establishes this for the stated record fields and stipulated facts.

Current-state checks prevent unauthorised release in the unconstrained-selection histories, where assessment errors and unnecessary blocking persist. Correcting those errors requires fixing route selection. The stale-state histories require a different correction because the supporting conditions change before commitment. Both the predicate-based and procedural implementations match the reference with the same requirements and several shared routines. The comparison tests whether the two enforce the specified conditions. It provides no measure of relative performance and cannot reveal defects common to the shared specification.

Two 40-unit proposals each fit a 70-unit cap when tested against the unused budget. Together they require 80 units, an example of locally valid decisions whose combined effects violate an invariant \parencite{bailis2014coordination}. The model checks each proposal against the budget left by earlier commitments. A distributed implementation must account for the delay and availability costs of coordinating that check. Partitioning capacity into exclusive spending rights could change the coordination needed at admission. Assessing that design requires accounting for transfers, revocation and recovery eligibility. The present results provide no latency or throughput estimate for either design.

Pooling recovery capacity can admit a loss that the available instruments cannot cover. Adding capacity to an instrument ineligible for that loss can leave the shortfall unresolved. Reservations and reassessments therefore need to preserve which instruments can cover each loss. In the live-control schedules, later capacity reductions make it infeasible to cover all admitted losses. An admission error calls for a corrected gate. If capacity deteriorates after admission, the bank must reassess outstanding obligations. Replenishment and exposure reduction each require authority for the action taken.

Keeping every request on hold prevents invalid commitments and can leave customers without resolution. Safety and liveness therefore require distinct arguments \parencite{alpern1985liveness}. The assigned owner, review deadline and escalation path specify who must act and when. A liveness argument would need to establish eventual resolution under stated conditions. Evaluation should record unresolved cases and time to an authorised disposition alongside invalid releases.

Respondents rated expected staff ease lower than usefulness and asked for clearer customer communication. Automation can leave operators responsible for rare failures while reducing their opportunity to practise handling them \parencite{bainbridge1983ironies}. The monitoring inequality specifies the time available to intervene. Measuring whether staff can meet it requires observing detection, handoff and stopping under workload. The survey captured expectations about the hold scenario before the authors refined the operational requirements.

The retention test checks whether the selected route, assessment, release decision, final routing label and payment reservation can be recovered from stored fields. Three are stored with commitment, one with selection and one with assessment, so the size of the score loss follows from which component is removed. Hash checking detects a change to retained bytes relative to the stored digest. Authentic capture and evidentiary weight require separate verification. A test of reviewability \parencite{cobbe2021reviewable} could use the paired payment histories to compare what reviewers can establish from a transaction log and from linked records. It would score whether reviewers identify the applicable mandate and applied rule, reach a supported diagnosis or correctly report that the available evidence leaves conformance indeterminate.

The finite designs test selected boundaries under authored mandates. Their 576 labelled selection contexts repeat 32 distinct behavioural combinations. The reference implements the research team's specification with separately coded decision logic and allocation enumeration. Literal cases supplement the reference. The deterministic executor uses authored evidence answers. Language-model reliability and strategic behaviour fall outside the evaluation. The insurance example illustrates routing. Extending the findings to claims settlement would require explicit entitlement rules and domain-specific loss and recovery inputs.

The simulated transition ends at internal booking and recording. If delivery is uncertain after dispatch, reconciliation must establish whether the payment took effect before the institution retries or releases the reservation. The FedNow customer credit transfer procedure describes a specific case in which the sender, after receiving a no-match status response, can resend a \texttt{pacs.008} with its original FedNow message identifier \parencite[\S~15.1(b)]{fednow2026procedures}. Each attempt's message identifier belongs in the delivery record linked to the bank's stable internal request identifier. Compensation after partial execution is a separate operation, as in sagas \parencite{garciamolina1987sagas}. A compensating payment requires its own authority and leaves the history of the original release intact. A service-level test should observe withdrawal, commitment, dispatch and external effect separately, using delayed updates, duplicate requests and lost acknowledgements while reservations are outstanding. The pertinent outcomes are invalid commitments, duplicate effects, premature release of reservations and unresolved holds. Durable recording and authentic approval also require faults that exercise their own enforcement paths.

\section{Conclusion}

The bank can admit payments against shared limits and preserve recovery feasibility when it checks and books each demand in the same ordered transition, under the stated fixed-resource assumptions. In the finite tests, eligible allocation excludes the false admissions produced by treating recovery capacity as interchangeable. Both implementations meet the same conformance criteria. The paired payment example shows why an admitted payment can still require a separate judgment about the handling route. The retention study establishes which of five specified facts remain available after changes to the record. Evaluating how reviewers diagnose failures and how payment operations resolves holds or uncertain deliveries requires further study.

\printbibliography

\swDeclarationsPage
\end{document}